\documentclass[12pt,preprint]{aastex}

\shorttitle{N$_2$H$^+$ Observations of Molecular Cloud Cores in Taurus}
\shortauthors{Tatematsu et al.}

\begin{document}


\title{N$_2$H$^+$ Observations of Molecular Cloud Cores in Taurus}


\author{Ken'ichi Tatematsu, Tomofumi Umemoto, 
Ryo Kandori, \altaffilmark{1} and Yutaro Sekimoto}
\affil{National Astronomical Observatory of Japan, 2-21-1 Osawa, Mitaka,
Tokyo 181-8588, Japan}
\email{k.tatematsu@nao.ac.jp, umemoto@hotaka.mtk.nao.ac.jp, 
kandori@alma.mtk.nao.ac.jp, sekimoto.yutaro@nao.ac.jp}


\altaffiltext{1}{The Graduate University for Advanced Studies (Sokendai), 
Department of Astronomical Science,
2-21-1 Osawa, Mitaka, Tokyo 181-8588, Japan}


\begin{abstract}
N$_2$H$^+$ observations of molecular cloud cores in Taurus with the Nobeyama
45 m radio telescope are reported.
We compare ``cores with young stars'' with ``cores 
without young stars''.
The differences in core radius, linewidth, and core mass are small.
Linewidth is dominated by thermal motions
in both cases.
N$_2$H$^+$ maps show that the intensity distribution
does not differ much between cores without stars and those with stars.
This is in contrast to the result previously obtained in H$^{13}$CO$^+$
toward Taurus molecular cloud cores.
Larger degree of depletion of H$^{13}$CO$^+$
in starless cores will be
one possible explanation for this difference.
We studied the physical state of molecular cloud cores
in terms of ``critical pressure'' for the surface (external) pressure.
There is no systematic difference between
starless cores and cores with stars
in this analysis. Both are
not far from the critical state for pressure equilibrium.
We suggest that molecular cloud cores in which thermal support is dominated
evolve toward star formation by keeping close to the critical state.
This result is in contrast with that obtained in the intermediate-mass
star forming region OMC-2/3, where molecular cloud cores evolve by decreasing
the critical pressure appreciably.
We investigate the radial distribution of the integrated intensity.
Cores with stars are found to have shallow ($-1.8$ to $-1.6$) power-law
density profiles.
\end{abstract}




\section{Introduction}

It is of great interest when and how the process 
of star formation takes place. \citet{nak98} theoretically studied 
the condition for the onset of star formation on the basis of the
virial theorem.  
For the surface (external) pressure
$P_s$, we can define the critical pressure $P_{cr}$, which is a function
of the effective sound speed and the mass of the molecular cloud core
\citep{spi68,nak98}. 
There are two equilibria if $P_{cr} > P_s$, whereas
there is no equilibrium if $P_{cr} < P_s$.
\citet{nak98} suggested that dissipation of turbulence in the core
will result in the decrease in $P_{cr}$, which will lead to
the onset of star formation because of the absence of equilibrium state.
\citet{aso00} observed the intermediate-mass star forming region OMC-2/3 in Orion by using the H$^{13}$CO$^+$ $J = 1\rightarrow0$ line.
They found
that cores without young stars tend to have $P_{cr} > P_s$ and 
that cores with young stars tend to have $P_{cr} < P_s$.
This fact suggests that dissipation of turbulence leads to
the onset of star formation in OMC-2/3.

It seems that molecular cloud cores 
in low-mass star forming regions 
like Taurus have lower surface (external) pressure than those in massive
or intermediate-mass star 
forming regions like Orion\citep{tat93}.  
The present study is aimed to investigate whether dissipation of turbulence
plays an important role also in low-mass star forming regions.
In low-mass star forming regions, the choice of the molecular-line tracers
is crucial.  It is known that molecular depletion
seems to be serious for CO, HCO$^+$, CCS, CS,
and their isotopomers in starless, cold molecular cloud cores 
\citep{cas99,aik01,ber01,ber02,zli02,taf02,lee03,she03}.
On the other hand, N-bearing molecules, NH$_3$ and N$_2$H$^+$
do not show strong evidence of depletion,
except for the very center of the cores 
\citep{ber01,ber02,cab02,taf02,lee03,shi03}.
Furthermore,
although H$^{13}$CO$^+$ is one of the popular tracers 
of the molecular cloud core,
it is known that the linewidth of the H$^{13}$CO$^+$ core
can be broaden by the molecular outflow, and this is
more serious in studying low-mass star forming regions \citep{aso00}.
We employ N$_2$H$^+$, which is less affected by depletion or molecular outflow.
N$_2$H$^+$ is found to trace the quiescent
molecular gas \citep{wom93,bac97}, 
and this choice would minimize the influence of
the molecular outflow.

The target molecular cloud cores are those observed by \citet{miz94} in H$^{13}$CO$^+$.
Molecular cloud cores associated with young stars are called
``cores with stars.''
Molecular cloud cores not associated with young stars
are called ``starless cores.''
Our samples contain three starless cores and four cores
with stars. We happen to detect one new starless core during the observations. 
The distances to cores in Taurus are assumed to be 140 pc. At this distance,
1 arcmin corresponds to 0.041 pc.

\section{Observations}

We carried out our observations by using the 45 m radio telescope
of Nobeyama Radio Observatory from 2003 January 7 to 9.  
The receiver front end was
the 25-element double-sideband focal-plane SIS receiver BEARS, which
has a beam separation of 41$\farcs$1. 
We observed molecular cloud cores in  
N$_2$H$^+$ $J = 1\rightarrow0$ at 93.1737767 GHz \citep{cas95}.
We mapped regions by shifting the telescope by half the beam separation,
so the final map consists of data on a 20$\farcs$55 spacing grid.
During the observations, one element of BEARS was not available due to
a technical problem, so we used 24 elements.
The half-power beamwidth of the element beams of
BEARS was estimated to be 17$\farcs$8$\pm$0$\farcs$4 at 93 GHz, which was close to
the data grid spacing.
The employed receiver back end was a digital autocorrelator.
We selected the spectral resolution to be 37.8 kHz (corresponding to $\sim$
0.12 km s$^{-1}$).
Spectra were obtained in the position-switching mode.
To compensate for the daily intensity scale variation,
we observed a local intensity peak at 
R.A. (J2000) = $5^h35^m16\fs0$,
Decl.(J2000) = $-5\degr19\arcmin26\arcsec$
near Orion KL with BEARS every day.
Furthermore, we observed 
this intensity peak using a single-sideband
SIS receiver S100 to calibrate the gain of each BEARS element
and to establish the absolute intensity scale.
The line intensity is reported in terms
of the corrected antenna temperature $T_A^*$ to be observed with the
S100 receiver in this paper.
The main-beam efficiency with S100 at 93 GHz is estimated to be 0.515,
by interpolating the measurements at 86 and 100 GHz by the observatory.
The telescope pointing was established by observing the SiO maser source 
NML Tau at 43 GHz every 1$-$1.5 hours during the observations.
The data were reduced by using the software package NewStar
of Nobeyama Radio Observatory and IDL of Research Systems, Inc.

Prior to our observations, 
sixty-three low-mass molecular cloud cores have been  
observed in N$_2$H$^+$ by \citet{cac02}, but with lower angular
resolution (54 arcsec).
We will refer to their results to secure our discussion
on the basis of our observations toward the eight cores
but with better angular resolution (18$\farcs$6 beam and 20$\farcs$55 grid).
Since the typical radius of the molecular cloud core is about
50 arcsec in the region, the improvement in spatial resolution 
helps us to derive
the physical parameters more precisely.




\section{Results}

Figure 1 shows that the distribution of the velocity-integrated intensity of
the main N$_2$H$^+$ $J = 1\rightarrow0$ 
component ($F_1$, $F$ = 2, 3$\rightarrow$1, 2)
toward the eight molecular cloud cores. 
The core associated with L1551 IRS5 (L1551) is 
2$-$3 times as intense as the other cores,
and we double the contour interval for this source
for clarity.
We use the main component rather than the
optically thinner components to obtain better signal-to-noise ratios.
For checking, we have also made the integrated intensity map
of the $F_1$, $F$ = 0, 1$\rightarrow$1, 2 component (not shown, 
the map quality is 
much worse), and have confirmed that the intensity peak position
is consistent in general.
As shown later, the optical depth is not very
large even for the main component ($\sim$ unity).
The positions of the intensity maxima of the cores 
are summarized in Table 1.
The core name with the prefix of Miz refers to 
the core number in \citet{miz94}.
Core Miz8b is newly found in our observations.
Core L1527 is associated with a Class 0 protostar, 
while cores Miz7, Miz8, and, L1551 are associated with Class I protostars.
These are ``cores with stars.''
Furthermore, a Class 0 protostar L1551 NE \citep{bar93} 
is located near the core boundary
of L1551, but its relationship to the N$_2$H$^+$ core is not clear in our map 
(see \citet{sai01,yok03} for a detailed study of this region in C$^{18}$O and 
CS).
Cores Miz1, Miz2, L1521F, and Miz8b are ``starless cores''.

The basic physical parameters of the cores 
are summarized in Table 2.
The HWHM core radius $R$ is measured as $\sqrt{S}/\pi$ ($S$ is the core area
S at the half maximum), and then corrected for the
telescope beam size. 
There is no remarkable difference in core radius 
between starless cores
($R$ = 0.035$\pm$0.004 pc) and core with stars ($R$ = 0.031$\pm$0.006 pc). 
The study of \citet{cac02} shows that cores with stars
are slightly larger than starless cores, but difference is not large.
Our result is in marked contrast with that of \citet{miz94} in 
H$^{13}$CO$^+$, in which cores with stars tend to be more compact.
Depletion of H$^{13}$CO$^+$ will be
a plausible explanation, because it is known that N$_2$H$^+$
is more robust for depletion than H$^{13}$CO$^+$ \citep{ber01,cab02,lee03}.
It is suggested that depletion becomes less prominent in cores
with stars because molecules evaporate from the dust grain
due to radiation and outflow from the newly formed star \citep{lan00}.
If some molecule is depleted in starless cores, we expect that
intensity tends to be weak in starless cores, and the radial
intensity profile is more flat-topped (because depletion is
more effective at the high-density core center), causing larger radius
in starless cores.
The tendency observed in H$^{13}$CO$^+$ can be explained if this molecule is 
depleted substantially only
in starless cores.

Next, we derive the line optical depth, linewidth, and mass.
We fit the observed spectrum by using the hyperfine spectrum model
consisting of multiple Gaussian components with line optical depth effect.
The intrinsic relative intensities of the hyperfine components
are taken from \citet{tin00}.
The free parameters are the excitation temperature $T_{ex}$,
the sum of optical depths of the hyperfine components $\tau_{TOT}$, systemic
velocity (radial velocity), and intrinsic linewidth (which is
not broadened due to line optical depth).
The details of the column density estimation are given in
\citet{cab02}.
We used the optically thick formula for the central 3$\times$3 
positions, and 
the optically thin formula for the weaker, outer part.
Figure 2 shows the results of the hyperfine fitting,
and Table 2 lists the physical parameters from the hyperfine fitting.
The excitation temperature at the intensity peak is
$T_{ex} = 6.2\pm2.3$ K for the seven cores excluding Miz1.
That for the composite spectrum from 
the central nine (3$\times$3) positions 
is $T_{ex} = 5.7\pm1.2$ K for the eight cores.
The optical depth of the main component ($F_1$, $F$ = 2, 3$\rightarrow$1, 2),
which is equal to 0.259 $\tau_{TOT}$,
is found to be moderate (1.54$\pm$0.78 for the intensity peak and
1.07$\pm$0.65 for the central nine positions).
The derived linewidth is then corrected for the frequency resolution
of the spectrometer.
Figure 3 plots the intrinsic linewidth against the angular distance
from the core center (impact parameter) $b$.
This corresponds to the Type 4 (single-tracer, single-cloud)
linewidth-size relation in \citet{goo98}.
The power-law index of the linewidth-size relation 
obtained by non-linear least-squares fitting is listed in Table 3. 
\citet{cac02} showed that molecular cloud cores 
have a variety of linewidth-size relation;
cores show positive, flat, and negative correlation of linewidth with
the impact parameter.
Our samples show that linewidth decreases or constant with increasing $b$ 
in general (Table 2).
The larger linewidth at the center in cores with stars 
could be due to the influence of protostellar
collapse \citep{zho94,cas02} and/or molecular outflow \citep{aso00},
although we selected the molecular line to minimize such effects.
To discuss the intrinsic core properties by eliminating this effect,
we use the average linewidth for $b > 40\arcsec$ in Miz8 and L1527 
and for $b > 80\arcsec$ in L1551.
The intrinsic N$_2$H$^+$ linewidth of the starless core is 
$\Delta v$ = 0.256$\pm$0.024 km s$^{-1}$,
while that of the core with star is 
$\Delta v$ = 0.309$\pm$0.070 km s$^{-1}$. 
So, there is no significant
difference.
By assuming a gas kinetic temperature of 10 K, we derive
the non-thermal linewidth $\Delta v_{NT}$ 
and the total linewidth $\Delta v_{TOT}$ \citep{ful92}.
$\Delta v_{NT}$ is estimated to be
0.224$\pm$0.030 km s$^{-1}$ and 0.284$\pm$0.082 km s$^{-1}$
for starless cores and cores with stars, respectively.
$\Delta v_{TOT}$
is derived to be 0.497$\pm$0.013 km s$^{-1}$ and 0.528$\pm$0.042 km s$^{-1}$
for starless cores and cores with stars, respectively.
The thermal linewidth $\Delta v_{T}$ for the mean molecular weight (2.33 $m_H$)
at 10 K is 0.443 km s$^{-1}$.
So, both starless cores and cores with stars are dominated by
the thermal support.
The effective pressure is 1.26$\pm$0.07 and 1.43$\pm$0.23
times as large as the thermal pressure for starless cores and
cores with stars, respectively.
Table 3 lists the H$_2$ column density $N$(H$_2$) obtained 
for the spectrum at the intensity peak, and for the
composite spectrum from the nine (3$\times$3) positions
centered on the intensity peak.
We derive the core mass $M$ by integrating the column density 
over the core.
We adopted the abundance of N$_2$H$^+$ to be
3.0$\times10^{-10}$ relative to H$_2$ \citep{cac02}.
The core mass $M$ is listed in Table 3. The average value is
$M$ = 1.30$\pm$0.63 $M_\sun$ for the starless cores,
and is 
$M$ = 1.59$\pm$0.97 $M_\sun$ for the cores with stars. 
These values are not much different, taking into account the uncertainty
in mass estimation (a factor of 2).

\section{Discussion}

We investigate whether the cores have equilibrium states or not
by following the discussion of \citet{nak98} 
(see \S 1 of the present paper 
for a brief summary).
The virial theorem is useful
not only for stable states but also for the unstable states not very far 
from the critical state \citep{nak98}.
For consistency, the formula in \citet{aso00} is used again,
$P_{cr} = 1/12\pi G^3 M^2 \times (5/3)^3 \times (9/4  C_{eff}^2)^4$,
where the effective sound speed $C_{eff}$ is 
$C_{eff}$ = ($\Delta v_{TOT}^2/$8 ln 2)$^{1/2}$.
Table 3 lists the critical pressure $P_{cr}$.
$P_{cr}$ is derived to be
(5.4$\pm$2.6)$\times10^{5}$
and (5.9$\pm$2.2)$\times10^{5}$
K cm$^{-3}$ for starless cores and cores with stars, respectively,
when we do not correct for the decrease in mass (see below).
When we correct for the decrease in mass due to protostellar collapse
and outflow for cores with stars,
the critical pressure $P_{cr}$ is derived
to be  (5.4$\pm$2.6)$\times10^{5}$
and (3.5$\pm$1.6)$\times10^{5}$
K cm$^{-3}$ for starless cores and cores with stars, respectively.
We estimate $P_s$ in Taurus to be of order 4$\times10^5$ K cm$^{-3}$ 
from the column density of low-density molecular gas and the coefficient
of the linewidth-size relation
\citep{tat93,oni02}.
This means that both starless cores and cores with stars are close to
the critical state $P_{cr} \sim P_s$ in Taurus.

To know whether turbulence dissipation leads to
the onset of star formation,
we need to know the physical condition prior to star formation.
The derived mass of the molecular cloud core is the current value.
This may have decreased from the original value,
because part of the original core mass has lost through star formation.
We estimate the decrease in the core mass due
to these star formation activities as follows.
Part of the core mass will be accreted onto the protostar, and part
of it will be swept up by the molecular outflow.
The high-velocity wind from the protostar will sweep up the ambient matter
(mostly the parent molecular cloud core, and possibly also the less dense
envelope), and this entrained mass together with the wind mass 
constitutes to the outflow mass.
The mass of the T Tauri star is typically 0.5 $M_\sun$.
The Class I protostar is supposed to have accreted more than half of the
final stellar mass \citep{bon96}.
The mass of the forming star is estimated to be 0.1-0.15 $M_\sun$
for IRAS 04169+2702 from the accretion luminosity estimate 
\citep{ohb97}.
The mass of the central star of L1551 IRS5 
is estimated to be 0.15$-$0.5 $M_\sun$
from the kinematics of the protoplanetary disk \citep{mom98}
and to be 0.7$-$3 $M_\sun$ from the stellar luminosity and disk mass 
\citep{sai96}.
The mass of the central star of IRAS 04368+2557 (L1527) is estimated
to be 0.1 $M_\sun$ from the kinematics of the protoplanetary disk and 
the accretion luminosity \citep{oha97}.
\citet{hog98} estimated the maximum central star mass to be
2.6 and 0.2 $M_\sun$ for L1551 IRS5 and IRAS 04368+2557 (L1527),
respectively, from the accretion luminosity.
IRAS 04166+2706 (Miz7), IRAS 04169+2702 (Miz8) and 
IRAS 04368+2557 (L1527) are known to accompany molecular outflows
\citep{bon96}.
The mass of molecular outflow associated with L1551 IRS5 is
0.3 $M_\sun$ \citep{lad85}.
\citet{hog98} estimated the outflow mass 
from $^{12}$CO (3$\rightarrow$2) to be
3.1 and 0.18 $M_\sun$ for L1551 IRS5 and IRAS 04368+2557 (L1527),
respectively.
Taking them into account, we estimate the mass lost from the
parent molecular cloud core is about 0.4, 0.4, 1.0, and 0.2 $M_\sun$ 
for IRAS 04166+2706 (Miz7), IRAS 04169+2702 (Miz8), L1551 IRS5, and
IRAS 04368+2557 (L1527), respectively.

Figure 4 plots
the critical pressure $P_{cr}$ 
against the core mass $M$.
There is no systematic difference between starless cores and cores with stars
in Taurus.
Regardless of starless cores or cores with stars, six
out of the eight molecular cloud cores are
located near the critical state of equilibrium.
Two cores, L1521F and L1551 are located slightly below $P_s$.
However, $P_s$ may have local variation and 
mass estimation will be uncertain by a factor of 2. 
It is hard to conclude that
these two cores are far from the critical state.
Although there is a possibility 
that the core in Taurus slightly dissipates
turbulence resulting in star formation and then the star formation activity 
increases the nonthermal linewidth,
the present observations do not provide us with the evidence
for the dissipation of turbulence.
Within the accuracy of the current study,
we conclude that 
both starless cores and cores with stars are close to the critical state.
\citet{mye83} obtained similar results
from NH$_3$
observations of low-mass molecular cloud cores;
cores are close to the critical state for equilibrium and stability
if the Doppler linewidth supports cores.
\citet{cac02} studied the ratio
of the core mass to the virial mass, and derived to be
$1.3\pm0.3$ and $1.4\pm0.3$ for starless cores and cores with stars,
respectively.  These values are almost identical: both cores
are close to virial equilibrium.
We suggest that the thermally-supported cores evolve keeping 
the critical state for equilibrium.
This result for the low-mass star forming region
is in contrast with that obtained in the intermediate-mass star forming
region OMC-2/3. In OMC-2/3, starless cores and cores with stars show clearly
different states.

We should revisit OMC-2/3 to see whether the results reported are
still correct even if we take into account the depletion
of H$^{13}$CO$^+$.
The molecular cloud cores in OMC-2/3 is warmer ($\sim$ 20 K, \citet{ces94}),
and show active star formation.
The sublimation temperature of the CO ice is 16 K in cloud cores
\citep{lan00}.
So, depletion will not be serious even for H$^{13}$CO$^+$.
\citet{aso00} have not taken into account the mass accreted onto the protostar
or swept by the outflow, 
but this correction will simply enhance
the observed difference between starless cores and cores with stars.
We conclude that the results obtained by \citet{aso00} is unchanged even
in the present context.

Lastly, we investigate the
radial density profile on the basis of the radial distribution of 
the integrated intensity
of the main component (Fig. 5).
We use here the integrated intensity, because it is
straightforward and more reliable than the hyperfine fitted column density
for outer regions.
The intensity profile is fitted with a power-law
$I \propto r^{p}$
convolved with the telescope beam.
The fitting results are listed in Table 3.
Miz1 and Miz2
may have (a hint of) central flattening.
For Miz1 and Miz2, the fitting result only for 
the outer part ($>$ 35 arcsec or $>$ 4800 AU) is shown
as broken lines in Figure 5.
The central flattening in cores in low-mass star forming regions 
was previously reported by
\citet{war99,and00,cac02} from the dust continuum observations and
N$_2$H$^+$ observations.
However, the integrated intensity profile in Miz1 and Miz2 
can also be fitted reasonably
with single (shallow) power laws (solid lines).
The data quality of the present observation is not enough to
distinguish these models.
The power-law index of the radial distribution is $p = -1.00\pm0.14$
(using the outer part for Miz1 and Miz2)
and $-0.72\pm0.07$ for starless cores and cores with stars,
respectively.
When we assume that the column density is proportional to
the integrated intensity, the power $p$ in the intensity profile is
related to the power $\alpha$ in the density profile $\rho \propto r^{\alpha}$
as $\alpha = p - 1$.  
L1521F was observed in N$_2$H$^+$ with the Berkeley-Illinois-Maryland 
Association (BIMA) millimeter array and with SCUBA on the James
Clerk Maxwell Telescope (JCMT) \citep{shi03}.  
These observations show the centrally peaked
intensity profile of L1521F (from their Figure 3, 
the BIMA intensity profile is 
fitted with $p \sim -1.3$
and the SCUBA intensity profile is fitted with 
$p \sim -1.0$ for $r >$ 10 arcsec).
These data do not show the central flattening for L1521F,
when we take into account their map resolution.
Some starless cores (Miz1 and Miz2) 
may have central flattening, while some starless cores may not.
On the other hand, there is no hint of
central flattening in cores with stars in our samples.
However, it is possible that central flattening was not observed
due to the limited spatial resolution ($\sim$ 18 arcsec or 2500AU) 
and data sampling
($\sim$ 21 arcsec or 2900 AU)
in our observations.
Our result shows that
cores with stars have shallow power-law density (or integrated intensity) 
distribution.
It is interesting that this is in contrast to the result 
in the lower spatial resolution 
study by \citet{cac02}, where cores with stars show steep integrated intensity
distribution.
There is a possibility that the shallow density profile in cores with stars 
can be results of the core collapse \citep{fos93} 
and/or core dispersal due to
the molecular outflow.
The present observations are probably not enough for disentangling 
these possibilities,
and further observations (high-resolution dust continuum map, 
near-infrared color-excess map, higher resolution N$_2$H$^+$ imaging, etc) 
are desirable.

\section{Summary} 

On the basis of N$_2$H$^+$ observations toward Taurus,
we have studied the physical properties of the molecular cloud core.
The core radius, linewidth, and intensity
distribution do not much differ between
starless cores and cores with stars.
This result is in contrast with that previously obtained in
H$^{13}$CO$^+$.
We suggest that depletion of H$^{13}$CO$^+$ causes this difference.
From the critical pressure analysis for Taurus cores,
there is no systematic difference between
starless cores and cores with stars. Both are
not far from the critical state for equilibrium.
We suggest that the starless cores which are almost thermally supported
evolve toward star formation by keeping close to the critical state.
This result is in contrast with that obtained in the intermediate-mass
star forming region OMC-2/3, where the molecular cloud core evolves
by dissipating turbulence largely.
The density profile is investigated from
the integrated intensity distribution in the cores. 
Cores with stars show shallow density profiles, $r^{-1.8}$ to $r^{-1.6}$.



\acknowledgments

K. T. is grateful to Takenori Nakano for comments on the draft 
and to Jeong-Eun Lee for discussion.

\clearpage


\begin{figure}
\plotone{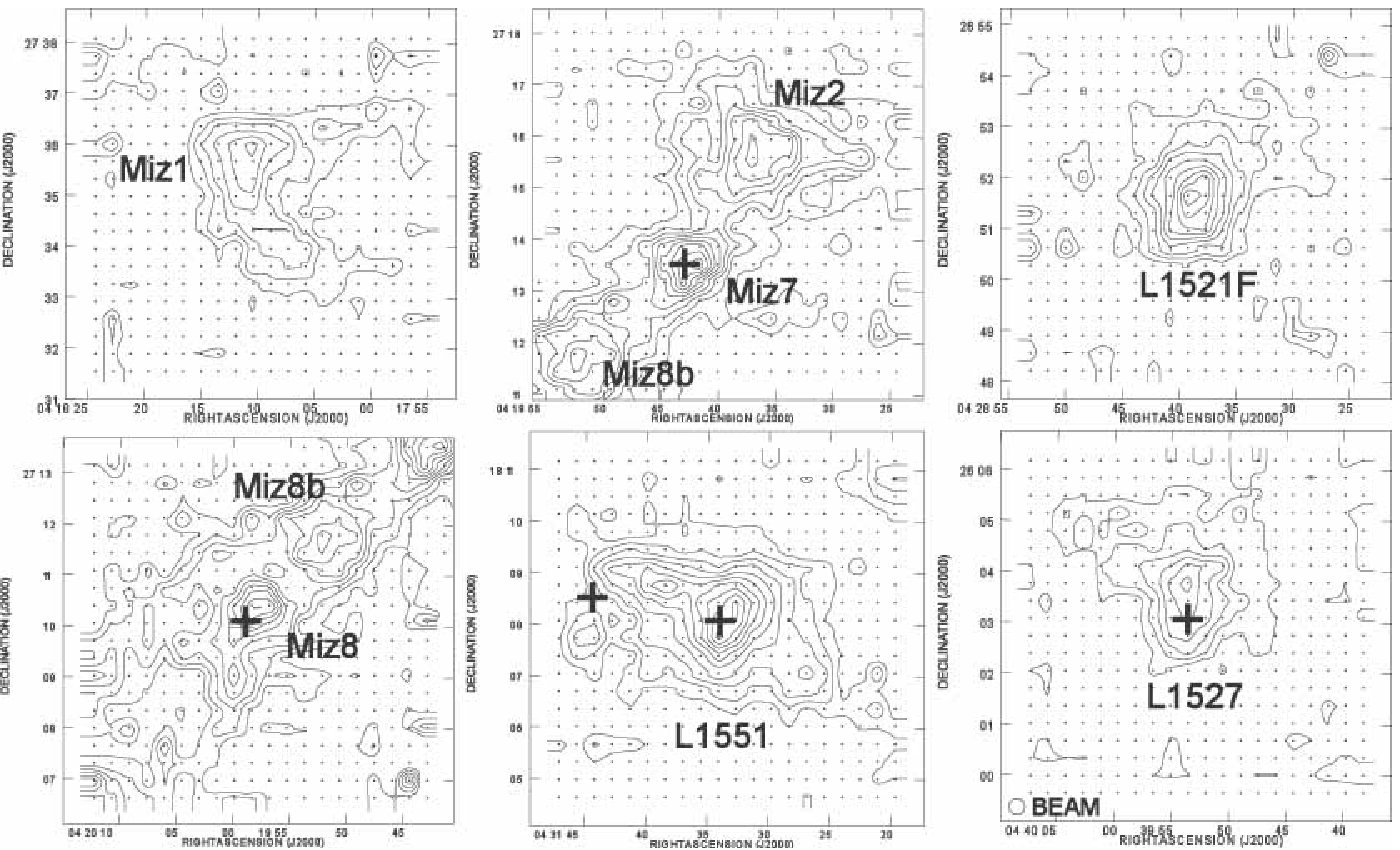}
\caption{Two-dimensional distribution of the velocity-integrated intensity of
the main N$_2$H$^+$ $J = 1\rightarrow0$
component ($F_1$, $F$ = 2, 3$\rightarrow$1, 2)
toward eight molecular cloud cores. 
The lowest contour level and level interval are 0.05 K km s$^{-1}$ except
for core L1551, for which they are 0.1 K km s$^{-1}$.
Small dots represent the observed positions.
Crosses represent the positions
of young stars. \label{fig1}}
\end{figure}

\clearpage 

\begin{figure}
\plotone{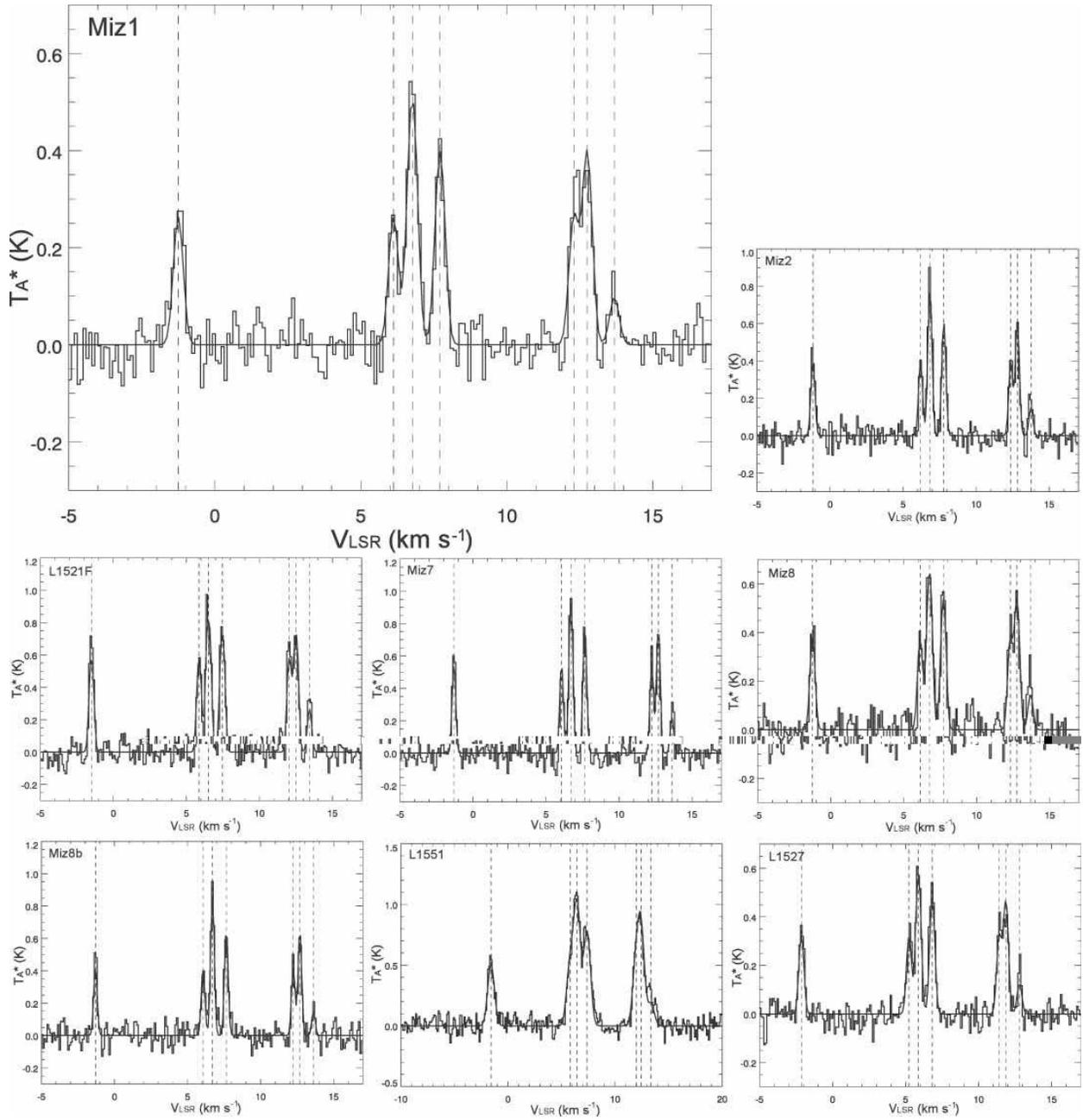}
\vspace{2cm}
\caption{Composite N$_2$H$^+$ spectra made by using
the central nine positions.
The result of the hyperfine fitting is also shown. \label{fig2}}
\end{figure}
\clearpage 

\begin{figure}
\plotone{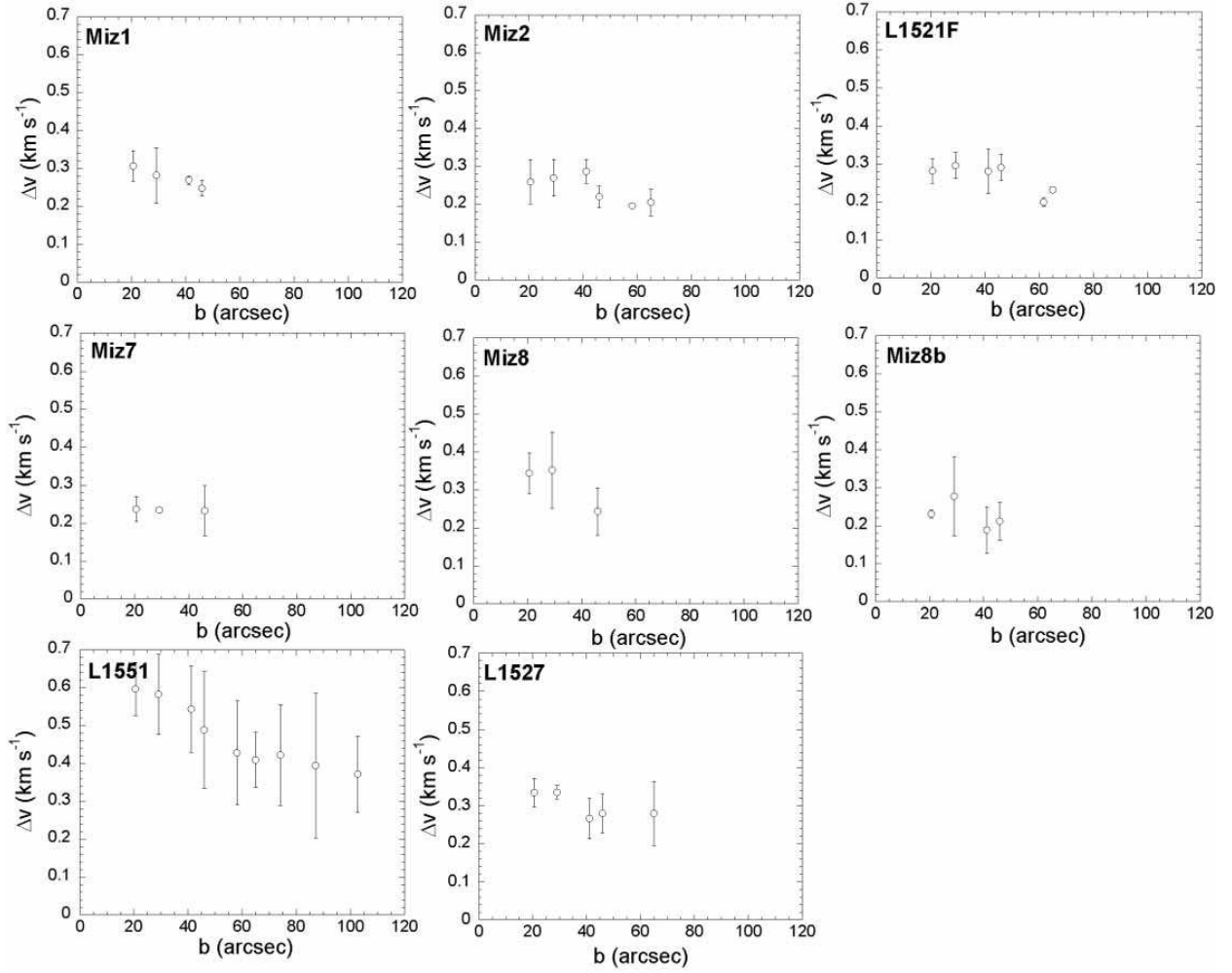}
\vspace{5cm}
\caption{The hyperfine fitted linewidth of
N$_2$H$^+$ is plotted against the angular distance from
the core center (impact parameter) $b$.
The open circle and error bar represent the average and
1$\sigma$ deviation of the linewidth at the same angular distance,
respectively.
\label{fig3}}
\end{figure}
\clearpage

\begin{figure}
\plotone{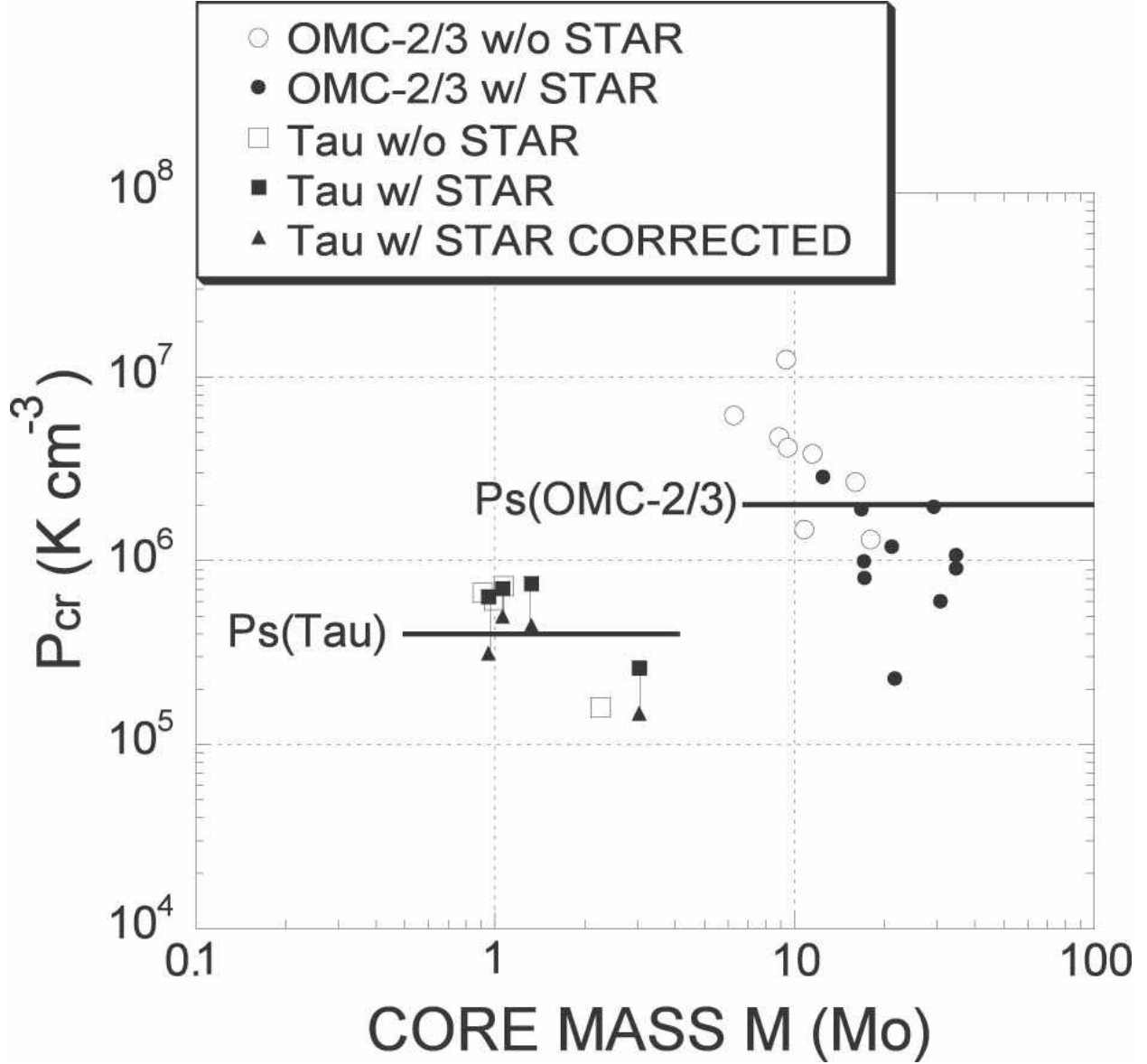}
\vspace{3cm}
\caption{The critical pressure $P_{cr}$ is plotted
against the core mass $M$.  For Taurus cores with stars,
$P_{cr}$ corrected for mass lost due to accretion onto the protostar
and due to the outflow is also shown as the filled triangle.
For reference, OMC-2/3 cores \citep{aso00} are also plotted.
The horizontal lines at $4\times10^5$ K cm$^{-3}$ 
and at $2\times10^6$ K cm$^{-3}$ represent
the estimated core surface (external) pressure $P_s$ for Taurus cores
and OMC-2/3 cores, respectively. \label{fig4}}
\end{figure}
\clearpage 

\begin{figure}
\plotone{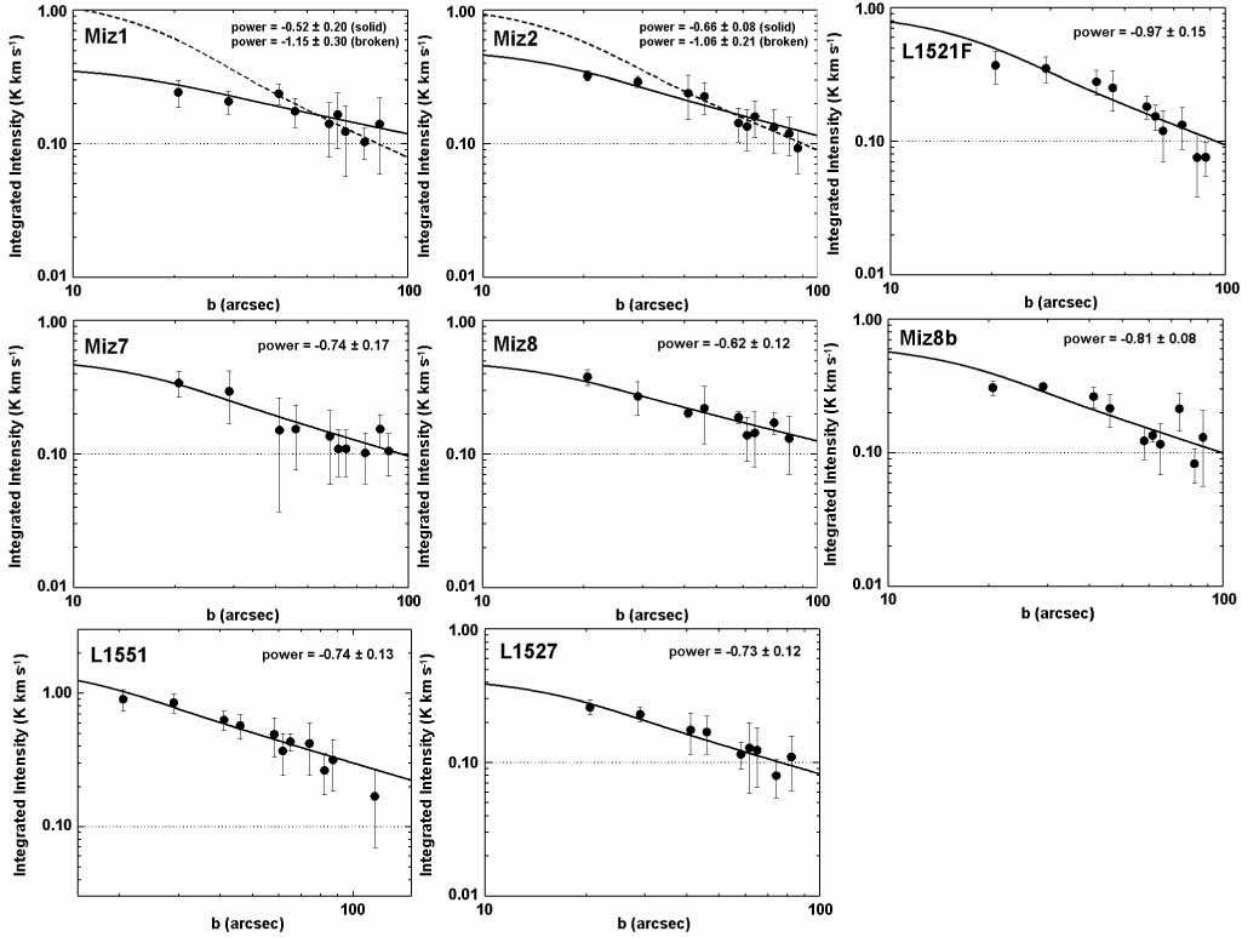}
\vspace{5cm}
\caption{The velocity-integrated intensity of
the main N$_2$H$^+$ $J = 1\rightarrow0$ 
component ($F_1$, $F$ = 2, 3$\rightarrow$1, 2)
plotted against the angular distance from
the core center (impact parameter) $b$.
The filled circle and error bar represent the average and
1$\sigma$ deviation at the same angular distance,
respectively.
The intensity profile is fitted with a power law
$I \propto b^{-p}$
convolved with the telescope beam.
For Miz1 and Miz2, the fitting results
by excluding two inner radius bins are also shown as broken lines.
\label{fig5}}
\end{figure}






\clearpage

\begin{deluxetable}{lccccc}
\tabletypesize{\scriptsize}
\tablecaption{Observed molecular cloud cores. \label{tbl-1}}
\tablewidth{0pt}
\tablehead{
\colhead{Field} & \colhead{R.A.(J2000)\tablenotemark{a}}   &
\colhead{Decl.(J2000)\tablenotemark{a}} 
& \colhead{Parent cloud name}
& IRAS$/$young star 
&Class
}
\startdata
Miz1    &4 18 10.3 &27 36 03 &L1495 &           &   \\
Miz2    &4 19 37.0 &27 15 35 &L1495 &           &   \\

L1521F  &4 28 39.3 &26 51 43 &L1521 &           &   \\
Miz7    &4 19 41.5 &27 13 32 &L1495 &04166+2706 &I  \\
Miz8    &4 19 58.2 &27 10 22 &L1495 &04169+2702 &I  \\
Miz8b   &4 19 52.2 &27 11 45 &L1495 &           &   \\
L1551   &4 31 32.5 &18 08 25 &L1551 &L1551 IRS5 &I  \\
        &          &         &      &L1551 NE   &0  \\
L1527   &4 39 53.2 &26 03 47 &L1527 &04368+2557 &0  \\

\enddata


\tablenotetext{a}{The maximum position of the
main N$_2$H$^+$ $J = 1\rightarrow0$ 
component ($F_1$, $F$ = 2, 3$\rightarrow$1, 2)
is listed.
Units of right ascension are hours, minutes, and seconds,
and units of declinations are degrees, arcminutes, and arcseconds.}

\end{deluxetable}

\clearpage

\begin{deluxetable}{lrcccc}
\tabletypesize{\scriptsize}
\tablecaption{Radius, velocity, linewidth, optical depth, 
and excitation temperature. \label{tbl-2}}
\tablewidth{0pt}
\tablehead{
\colhead{Core} 
&\colhead{$R$ (pc)} & 
\colhead{$V_{LSR}$ (km s$^{-1}$)\tablenotemark{b}}   & 
\colhead{$\Delta v$ (km s$^{-1}$)\tablenotemark{c}}   & 
\colhead{$\tau_{TOT}$\tablenotemark{b}}   & 
\colhead{$T_{ex}$ (K)\tablenotemark{b}}   

}
\startdata
Miz1  &0.038 &6.75$\pm$0.01	&0.28$\pm$0.04  &---\tablenotemark{d}&---\tablenotemark{d}\\
      &      &(6.76$\pm$0.01)	&               &(2.81$\pm$1.07)&(4.86$\pm$0.62)\\
Miz2  &0.036 &6.83$\pm$0.01	&0.24$\pm$0.05  &4.00$\pm$1.33 &6.23$\pm$0.83\\
      &      &(6.84$\pm$0.01)	&               &(2.73$\pm$0.86)&(5.99$\pm$0.80)\\
L1521F  &0.029 &6.50$\pm$0.01	&0.28$\pm$0.04  &9.72$\pm$1.79 &5.07$\pm$0.23\\
      &      &(6.51$\pm$0.01)	&               &(9.63$\pm$0.87)&(4.65$\pm$0.09)\\
Miz7  &0.024 &6.72$\pm$0.01	&0.24$\pm$0.05  &7.64$\pm$1.33 &5.78$\pm$0.31\\
      &      &(6.70$\pm$0.01)	&               &(5.03$\pm$0.72)&(5.47$\pm$0.26\\
Miz8  &0.029 &6.75$\pm$0.01	&0.33$\pm$0.05  &7.92$\pm$1.78 &5.02$\pm$0.29\\
      &      &(6.77$\pm$0.01)	&               &(4.54$\pm$1.06)&(4.78$\pm$0.33)\\
Miz8b &0.036 &6.73$\pm$0.01	&0.23$\pm$0.06  &3.96$\pm$1.79 &5.91$\pm$1.03\\
      &      &(6.71$\pm$0.01)	&               &(1.89$\pm$0.82)&(7.40$\pm$1.67)\\
L1551 &0.038 &6.41$\pm$0.02	&0.40$\pm$0.10  &1.06$\pm$0.79 &11.20$\pm$5.53\\
      &      &(6.43$\pm$0.01)	&               &(2.19$\pm$0.42)&(7.60$\pm$0.73)\\
L1527 &0.032 &5.85$\pm$0.01	&0.27$\pm$0.07  &7.43$\pm$1.94 &4.20$\pm$0.22\\
      &      &(5.88$\pm$0.01)	&               &(4.21$\pm$0.88)&(4.60$\pm$0.28)\\

\enddata


\tablenotetext{b}{Upper row shows the value at the intensity peak, 
and the lower row in parentheses is the value from the composite 
spectrum of central nine positions.
The error represents 1$\sigma$ error in the hyperfine fitting.}
\tablenotetext{c}{Intrinsic 
N$_2$H$^+$ linewidth.
For Miz8, L1551, and L1527, the values derived for the core edge ($b >$ 40$\arcsec$,
80$\arcsec$, and 40$\arcsec$, respectively) are used
to minimize the effect of the protostellar collapse and molecular outflow.
For the remaining sources, the average values over the cores are used.}
\tablenotetext{d}{Very large uncertainty.}

\end{deluxetable}

\clearpage

\begin{deluxetable}{lccccc}
\tabletypesize{\scriptsize}
\tablecaption{Column density, mass, critical pressure, 
radial velocity distribution,
and radial intensity distribution. \label{tbl-3}}
\tablewidth{0pt}
\tablehead{
\colhead{Core}  
& \colhead{$N$(H$_2$) (cm$^{-3}$)\tablenotemark{e}}
& \colhead{$M$ ($M_\sun$)\tablenotemark{f}}  
& \colhead{$P_{cr}$ ($10^5$ K cm$^{-3}$)\tablenotemark{g}} 
& \colhead{Power index of linewidth\tablenotemark{h}} 
& \colhead{Power index of intensity distribution\tablenotemark{i}} 

}
\startdata
Miz1  &---\tablenotemark{j}      &1.1 &7.3     &-0.23$\pm$0.18 (C.C.=0.88) &$-1.2\pm0.3$\tablenotemark{k}\\
      &($1.0\pm0.4\times10^{22}$)&    &        &             &                             \\
Miz2  &$1.3\pm0.5\times10^{22}$ &1.0 &6.1     &-0.42$\pm$0.14 (C.C.=0.79)  &$-1.1\pm0.2$\tablenotemark{k}\\
      &($1.1\pm0.4\times10^{22}$)&    &       &              &                             \\
L1521F&$3.0\pm0.6\times10^{22}$ &2.2 &1.6     &-0.17$\pm$0.08 (C.C.=0.48)  &$-1.0\pm0.2$                 \\
      &($2.8\pm0.3\times10^{22}$)&    &        &             &                             \\
Miz7  &$2.2\pm0.4\times10^{22}$ &1.0 &3.1$-$6.4 &-0.02$\pm$0.34 (C.C.=0.98) &$-0.7\pm0.2$               \\
      &($1.4\pm0.2\times10^{22}$)&    &        &              &                             \\
Miz8  &$2.6\pm0.6\times10^{22}$ &1.3 &4.4$-$7.5 &-0.41$\pm$0.36 (C.C.=0.92) &$-0.6\pm0.1$               \\
      &($1.7\pm0.4\times10^{22}$)&    &        &              &                             \\
Miz8b &$1.3\pm0.6\times10^{22}$ &0.9 &6.7     &-0.14$\pm$0.25 (C.C.=0.68)   &$-0.8\pm0.1$                 \\
      &($0.9\pm0.4\times10^{22}$)&    &       &               &                             \\
L1551 &$3.2\pm2.7\times10^{22}$ &3.0 &1.5$-$2.6 &-0.31$\pm$0.13 (C.C.=0.97) &$-0.8\pm0.1$               \\
      &($3.0\pm0.6\times10^{22}$)&    &        &              &                             \\
L1527 &$2.2\pm0.6\times10^{22}$ &1.1 &5.0$-$7.1 &-0.23$\pm$0.19 (C.C.=0.79) &$-0.7\pm0.1$               \\
      &($1.4\pm0.3\times10^{22}$)&    &        &              &              \\

\enddata


\tablenotetext{e}{Upper row shows the value at the intensity peak, 
and the lower row in parentheses is the value from the composite 
spectrum of central nine positions.
The error represents 1$\sigma$ error in the hyperfine fitting.}
\tablenotetext{f}{Mass lost due to the forming star and the associated molecular outflow is not taken into account.}
\tablenotetext{g}{For cores with stars, the ``range'' represents
the values uncorrected (upper bound) and corrected (lower bound) for the
mass loss due to accretion onto the star and due to the molecular outflow.}
\tablenotetext{h}{Power-law index $q$ for the linewidth-size relation  (Fig. 3)
estimated by using the non-linear least-squares fitting in the form $\Delta v \propto b^q$. 
The correlation coefficient
is listed in parentheses.}
\tablenotetext{i}{Power-law index from the fit to the radial integrated
intensity distribution of the main component (Fig. 5).}
\tablenotetext{j}{Very large uncertainty.}
\tablenotetext{k}{Central part is excluded for fitting.}

\end{deluxetable}



\end{document}